\documentclass[prl,twocolumn,showpacs,preprintnumbers,amsmath,amssymb]{revtex4}

\usepackage{graphicx}
\usepackage{dcolumn}
\usepackage{bm}

\begin{document}

\title{Vortex-Antivortex Lattices in Superconducting Films \\ with Magnetic Pinning Arrays}

\author{M. V. Milo\v{s}evi\'{c}}
\author{F. M. Peeters}
\email{francois.peeters@ua.ac.be}

\affiliation{Departement Fysica, Universiteit Antwerpen (Campus Drie Eiken), \\
Universiteitsplein 1, B-2610 Antwerpen, Belgium}

\date{\today}

\begin{abstract}
Novel vortex structures are found when a thin superconducting film
(SC) is covered with a lattice of out-of-plane magnetized magnetic
dots (MDs). The stray magnetic field of the dots confines the
vortices to the MD regions, surrounded by antivortices which
``crystallize'' into regular lattices. First and second order
transitions are found as magnetic array is made sparser or
MD-magnetization larger. For sparse MD-arrays {\it fractional}
vortex-antivortex states are formed, where the crystal-symmetry is
combined with a {\it non-uniform} ``charge'' distribution. We
demonstrate that due to the (anti)vortices and the supercurrents
induced by the MDs, the critical current of the sample actually
increases if exposed to a homogeneous external magnetic field,
contrary to conventional SC behavior.
\end{abstract}


\maketitle

The physics of vortex-antivortex pairs in superconductors and
superfluids has been of general interest for a long time. For
instance, such pairs are predicted to exist in thin
superconducting films at finite temperatures due to thermal
fluctuations \cite{kosterlitz}. Entropy considerations show that
above the sharply defined Kosterlitz-Thouless transition
temperature $T_{KT}$, these vortex pairs start to unbind, causing
the appearance of a finite resistance. Recently it was found that
symmetry-induced antivortices can be formed in mesoscopic
superconducting polygons \cite{chibotaru} in a certain
parameter-range, such that the vortex-antivortex configuration
complies with the geometry of the polygon. In our recent work, we
studied the vortex structure of a superconducting film with a
single out-of-plane magnetized dot on top \cite{misko}. The total
flux penetrating the superconductor equals zero, and vortices
cannot form in isolation; vortices and antivortices nucleate in
pairs. A shell vortex structure was observed, with a vortex
nucleus surrounded by an antivortex core (the so-called
``vortex-molecule'') with size-magnetization controlled magic
numbers. These vortex configurations resemble the ones of electron
dimples on the surface of liquid helium, electrons in quantum
dots, colloidal suspensions and dusty particles in complex plasmas
\cite{bedanov}.

In the present article, we report further consequences of this
superconducting Wigner crystallization, in case when a regular
{\it array} of magnetic particles is deposited on the
superconducting film. Modern advances in microfabrication and
characterization techniques~\cite{schuller} have allowed an
experimental realization of such SC/ferromagnet(FM) hybrid
systems. Arrays of magnetic particles are potential devices for
applying well-defined local magnetic fields, which modulate the
order parameter in an underlying superconductor. Refs.
\cite{schuller2,vanbael} (and references therein) have explored a
plethora of physical effects, including matching effects with
ordered pinning arrays, where additional pinning contributions
arise due to the magnetic nature of the pinning centers.

\begin{figure}[b]
\includegraphics[height=3.8cm]{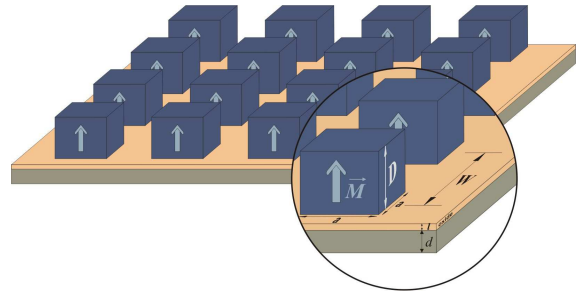}
\caption{\label{fig:fig1}Oblique view of the superconducting film
and oxide layer (with thicknesses $d$ and $l$, respectively)
underneath a regular array (with period $W$) of cubic magnetic
dots.}
\end{figure}

Here, we investigate the superconducting state of a thin SC film
with a square array of submicron cubic magnetic dots with
perpendicular magnetization (Fig. 1). To ensure that MDs and SC
are not electronically coupled, we assume a thin layer of
insulating oxide between them. We consider cubic MDs, although
most of the previous experimental work was done on thin FM
structures. Making the magnetic dots thicker facilitates their
magnetizing in the out-of-plane direction and eliminates the
extreme peak structure in the stray field profile close to the dot
edge. The general physical behavior of the SC drawn out in this
article is immune to the MD-thickness.

The energy difference between the superconducting and the normal
state, in units of $H_{c}^{2}\big/4\pi$, is
\begin{eqnarray}
\Delta \mathcal{G}_{s/n}&=&\int \left[
-|\Psi|^{2}+\frac{1}{2}|\Psi|^{4}+\frac{1}{2}|(-i\nabla-{\bf A})
\Psi|^{2} \right. \nonumber \\& &\left. +\kappa^{2}({\bf H}-{\bf
H}_{0})^{2} \right]dV, \label{freeen}
\end{eqnarray}
where ${\bf H}_{0}$ denotes the applied magnetic field (for
example, ${\bf H}_{0}={\bf H}_{md}$ if magnetic dots are the only
field source). Eq. (\ref{freeen}) is given in dimensionless form,
where all distances are measured in units of the coherence length
$\xi$, the vector potential ${\bf A}$ in $c\hbar /2e\xi$, the
magnetic field ${\bf H}$ in $H_{c2}=c\hbar /2e\xi ^{2}$, and the
order parameter $\Psi$ in $\sqrt{-\alpha /\beta}$ with $\alpha $,
$\beta $ being the GL coefficients. The minimization of Eq.
(\ref{freeen}) leads to well known GL equations which for thin
superconductors $(d<\xi ,\lambda)$ may be averaged over the SC
thickness. We solve these two coupled equations, following a
numerical approach proposed by Schweigert {\it et al.} (see Ref.
\cite{schweigert1} and references therein) on a uniform Cartesian
grid with typically 10 points/$\xi$ in each direction. In the
present case, we took for the simulation region a rectangle
$W_{x}\times W_{y}$, where $W_{x}=W_{y}=16W$ (i.e. we simulate $16
\times 16$ supercell, see Fig. 1). Periodicity of the SC and the
MD-lattice is included by applying periodic boundary conditions
for ${\bf A}$ and $\Psi $ in the form ${\bf A}({\bf r}+{\bf
b}_{i})={\bf A}({\bf r})+{\nabla}\eta _{i}({\bf r})$, and $\Psi
({\bf r}+{\bf b}_{i})=\Psi \exp(2\pi i\eta _{i}({\bf r})/\Phi
_{0})$~\cite{doria}, where ${\bf b}_{i=x,y}$ are the supercell
lattice vectors, and $\eta _{i}$ is the gauge potential. These
boundary conditions mean that ${\bf A}$, $\Psi $ are invariant
under lattice translations combined with {\it specific gauge
transformations}. Since the vector potential of a regular array of
magnetic dots is periodic by itself, we choose
$\eta_{x}=\eta_{y}=0$. If the sample is exposed to an additional
homogeneous perpendicular magnetic field ${\bf H}_{ext}$ (${\bf
H}_{0}={\bf H}_{md}+{\bf H}_{ext}$) we use the Landau gauge ${\bf
A}_{ext}=H_{ext}x{\bf e}_{y}$ for the external vector potential
and $\eta _{x}=H_{ext}W_{x}y$ while $\eta _{y}=0$. Note that
values of $H_{ext}$ may not be chosen freely and must fulfill the
{\it flux quantization per supercell} requirement following from
the virial theorem \cite{doria}.

To explore the superconducting state, we start from different
(randomly generated) initial configurations, increase/decrease
slowly (``sweep up/down'') the magnetization of the MDs $M$ and
let the vortex-configuration-solution relax to a steady-state one
(in principle, it may be metastable). For given $M$, we
recalculate the vortex structure of the film starting from: (i)
the previously found configuration during the sweep, (ii) Meissner
state $(\Psi\approx 1)$ or (iii) the normal state ($\Psi\approx 0$
in the whole sample) as initial condition. By comparing the
energies of all found vortex states we determine the ground state
configuration. The obtained $M-W$ equilibrium vortex phase diagram
is shown in Fig. 2 for the case of a SC film with thickness
$d=0.2\xi$ (this is, for example, an adequate value for a $50$nm
Pb film at $T/T_{c}=0.97$ \cite{vanbael}) and GL parameter
$\kappa=1.2$ (approximately corresponding to the experimental
values found for Pb, Nb, or Al films), covered with an oxide layer
(thickness ${\it l}=0.1\xi$) and an array of magnetic cubes with
$a=D=2\xi$ (see Fig. 1).

\begin{figure}[t]
\includegraphics[height=7.4cm]{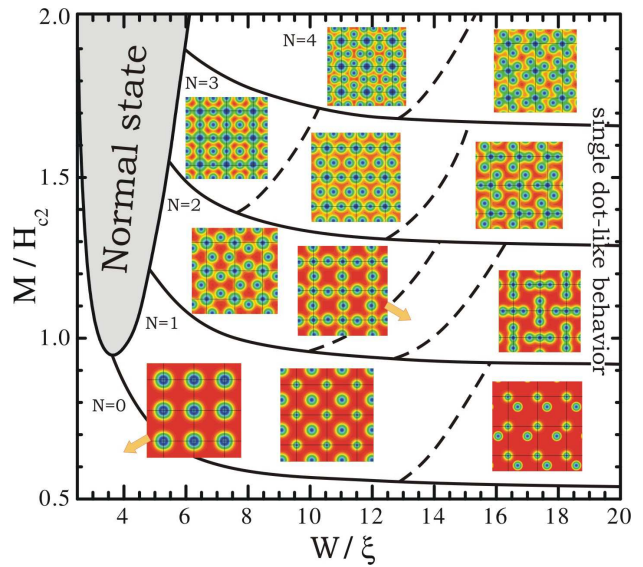}
\caption{\label{fig:fig2}Dependence of the vortex-antivortex
configurations on the magnetization ($M$) and period of the
magnetic dot (MD) lattice ($W$). Solid lines denote transitions
between states with different number of vortex-antivortex pairs
per unit cell ($N$), while dashed lines indicate second order
configurational transformations for fixed $N$. The vortex
structure is illustrated by the Cooper-pair density contourplots
as insets (blue/red - low/high density), where the thin lines
outline the unit cells of the MD lattice.}
\end{figure}
For small distance between MDs, the positive stray field under
each magnetic dot is compensated by the negative fields of the
neighboring dots, which decreases the amplitude of the magnetic
field modulation seen by the SC (see Fig. 3). Note that the total
flux through the SC is always zero. Since the demagnetizing factor
of an infinite magnetic film is unity, for $W=a$ the magnetic
field equals zero everywhere and the SC state exists for arbitrary
value of $M$. Hence, the magnetization value at which the S/N
transition occurs decreases with increasing distance between the
dots. In this region an unusual phenomenon occurs: increased
strength (magnetization) of the dots drives the SC directly to the
normal state, although the appearance of vortex-antivortex pairs
is expected \cite{misko}. Namely, even if the critical conditions
for their nucleation are achieved, there is not enough space for
stabilizing antivortices in the narrow negative field areas. At
the same time, the magnetic field under, and especially between
the MDs is so large (see $W=3\xi$ result in Fig. 3) that the SC
state is suppressed globally.

When increasing $W$, the compensation effect diminishes, the field
under MDs increases, suppressing locally the SC-state. Between the
dots, negative stray field becomes lower (spread over wider area)
and superconductivity survives. For small magnetization of the
dots, the magnetic field is only able to suppress the
superconducting order parameter under the edges of the MDs, where
the induced currents are maximal \cite{misko}. Such MDs act as
pinning centra for external vortices (e.g. resulting from the
application of an external homogeneous magnetic field). This
situation is very similar to the SC film perforated with a lattice
of antidots.
\begin{figure}[t]
\includegraphics[height=5.6cm]{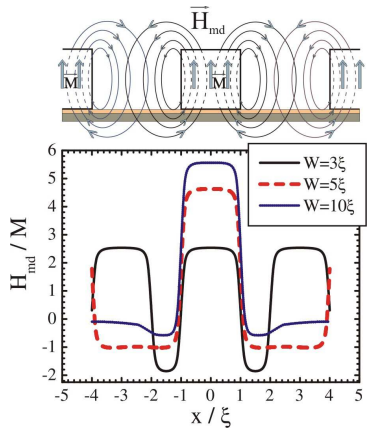}
\caption{\label{fig:fig3} Profile of the MD-magnetic field across
the sample for different values of the period of the MD-lattice.}
\end{figure}
By increasing the magnetization, the negative flux between the MDs
(as well as the positive one under the dots) increases, and the
magnetic field lines can ``join'' into an antivortex (vortex). The
critical magnetization and period $W$ for the first nucleated
vortex-antivortex pair per MD (in the ground state) are denoted in
the lower part of Fig. 2 by a solid line. Using magnetostatic
calculations, we determined that along this line, the positive
flux under each MD is approximately constant, and equals
$\Phi^{+}_{0 \mapsto 1}/\Phi_{0}=1.291 \pm 0.012$. We found that
the additional positive magnetic flux necessary for the nucleation
of the next vortex-antivortex pairs is quantized:
$\Delta\Phi^{+}/\Phi_{0}=1.07 \pm 0.01$. These values of threshold
fluxes weakly depend on the parameters of the SC, but strongly on
the properties of the stray magnetic field determined by the
geometrical dimensions of the MDs (i.e. $\Phi^{+}_{0 \mapsto 1}$
increases significantly with the size of the dots $a$).

For dense lattices, the antivortices are compressed into narrow
interstitial channels, forced to form regular and consequently
rigid lattices, where now antivortices are ``shared'' by the
neighboring MDs. The rigidness, together with the uniform
distribution of vortices and antivortices, makes these ordered
vortex structures resemble ionic crystals. In addition, one finds
significant similarities in the physical mechanisms of
crystallization. The somewhat simplified theory of cohesion in the
ionic (and molecular) crystals assumes that the cohesive energy is
entirely given by the potential energy of classical particles
localized at equilibrium positions. Because the particles in ionic
crystals are electrically charged ions, the main term in the
interaction energy is the interionic Coulomb interaction. The
other contribution comes from the strong short-range core-core
repulsion due to the Pauli principle, without which the crystal
would collapse. Analogy with our system follows from the present
relations - namely, vortices and antivortices interact analogously
to ions, except for the absent core-core repulsion, necessary for
crystallization. This stabilizing factor is brought in our sample
by the presence of magnetic dots, which effectively keep the
vortices and surrounding antivortices apart. Therefore, each
antivortex interacts with a magnetic dot coupled with the vortex
underneath through a Lennard-Jones-like potential, forming {\it
superconducting ion-pairs}, which when brought closer together
form a two-dimensional ionic crystal. This fascinating parallel is
best illustrated by the vortex-antivortex $N=1$ lattice from Fig.
2, which corresponds to the ion-configuration on the surface of a
NaCl crystal (and many other salts and oxides, e.g. AgBr, PbS,
FeO, etc.).

In $N=2$ crystal (each dot creates a double vortex and two
antivortices), antivortex dimers are shared between the
neighboring dots in such a way that each MD is surrounded by $4$
antivortices arranged in a cross. For small $W$, the adjacent
crosses are tilted with respect to each other. The tilt angle
changes with $W$ (or magnetization $M$, see dashed lines in Fig.
2) and the configuration transforms through a {\it second-order
phase transition} to square symmetry (tilt angle zero). This
bipartite crystal now consists of two sublattices (of vortices and
antivortices), where sites belonging to one lattice are connected
{\it only} to the sites of another (see, e.g., the surface of the
ReO$_{3}$ crystal). In $N=3$ lattice, the orientational degree of
freedom is lost, since antivortices crystallize in a perfect
square lattice. With increasing distance between the MDs, the
crystallization mechanism becomes more influenced by the
inter-(anti)vortex interaction than the imposed symmetry, and the
square lattice gradually transforms into a hexagonal one.

However, if the MDs are set further apart, the antivortex lattice
bonds gradually break, leading to oriented clusters (rather than a
crystal) of vortex-antivortex molecules around each MD, like in
the case of a single magnetic dot on top of a SC \cite{misko}. For
example, in $N=1$ state, the antivortices are no longer in the
central interstitial position, but are bound to a particular MD.
Their relative position is such that it maximizes the distance
between them. The $N=3$ molecules have a specific orientational
order in which molecules under adjacent columns of MDs are rotated
by $60^{\circ}$, due to the repulsion of neighboring
antivortex-trimers. Note that the distribution of antivortices in
this case does not obey the symmetry of the magnetic potential
(cubic MDs). This leads to an interesting scenario, where some of
the vortex-antivortex pairs may annihilate in order to preserve
the square symmetry of the vortex state at the expense of the
energy, allowing the vortex configuration to crystallize again. In
such manner, {\it fractional} states are formed, where some MDs
``share'' a vortex-antivortex pair (the number of pairs per dot
$N$ becomes a rational number). In Fig. 4 we show the Cooper-pair
density plots of two typical fractional states for the square
magnetic lattice $N=2^{1}/_{2}$ and $N=2^{3}/_{4}$. Here two
species of vortices are present (doubly and triply charged),
causing the adequate rearrangement of singly charged interstitial
antivortices. Although with higher energy than the ground-state
(for example, $\Delta E \approx 278kT_{c}$ for 50nm Nb or Pb films
at temperatures far from $T_{c}$), these states are metastable and
experimentally observable.
\begin{figure}[t]
\includegraphics[height=4.8cm]{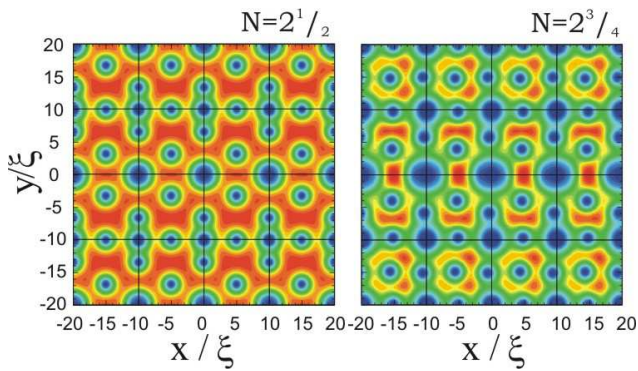}
\caption{\label{fig:fig4}The $|\Psi|^{2}$ contourplots of typical
{\it fractional} vortex-antivortex crystals for the $4 \times 4$
unit cell simulation region.}
\end{figure}

The predicted new vortex configurations can be observed
experimentally by using e.g. scanning probe techniques like Hall
and Magnetic Force Microscopy. These vortex-antivortex structures
will strongly influence the pinning properties and the SC phase
diagram. If one adds a homogeneous external magnetic field, such
that the number of additional flux quanta matches the number of
antivortices at the interstitial sites, one expects that
annihilation occurs, resulting in a well known matching vortex
configuration with all vortices pinned by the MDs. This leads to a
peak in the critical current, as a function of the applied
external field. This mechanism explains the recently observed
phenomenon of magnetic-field-induced
superconductivity~\cite{lange}. In order to verify this, we
exposed our sample with $W/\xi=6.25$ to a homogeneous magnetic
field corresponding to the first matching field (one vortex per
unit cell) and changed gradually the magnetization $M$, starting
each time from the normal state. Then we apply current in the
$x$-direction as $A_{cx}=const.$ (now ${\bf A}_{0}={\bf
A}_{md}+{\bf A}_{ext}+{\bf A}_{c}$) which does not interfere with
our boundary conditions. When the critical value of $A_{cx}$ is
exceeded, the motion of (anti)vortices can no longer be prevented
and superconductivity is destroyed. The results of our
calculations for the critical current $j_{c}$ as a function of the
magnetization of the MDs are shown in Fig. 5 for the case with and
without applied first matching field. If no external field is
present (dark dots in Fig. 5), higher magnetization $M$ induces
larger screening currents and $j_{c}$ monotonously decreases. The
appearance of vortex-antivortex pairs decreases the total current
in the sample through the phase-change contribution to the
current. This leads to somewhat enhanced critical current which
decreases further with magnetization and tends to zero. On the
contrary, in the case of the first matching field (open dots in
Fig. 5) the critical current equals zero if no pinning is present.
With increasing magnetization, the antivortex-like currents
\cite{misko} are increased, compensating the current of external
vortices pinned by the dots.
\begin{figure}[t]
\includegraphics[height=5cm]{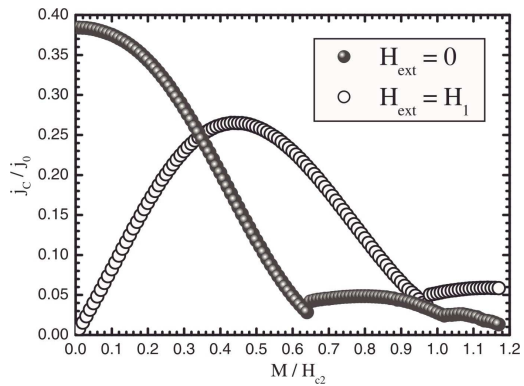}
\caption{\label{fig:fig5} Critical current (in units of
$j_{0}=cH_{c2}\xi \big/ 4\pi\lambda^{2}$) versus the magnetization
of the magnetic dot lattice ($a/\xi=2$, $W/\xi=6.25$) in the case
of no applied external magnetic field $H_{ext}$ (dark dots) and
for the first matching field (open dots).}
\end{figure}
For $M/H_{c2}=0.418$ maximal compensation is reached, resulting in
the maximal critical current. With further increased
magnetization, the qualitative behavior of $j_{c}$ is similar to
the $H_{ext}=0$ case. Nevertheless, if an external magnetic field
is present, the critical current of the sample for given $M$ is
found to be actually {\it higher}. This demonstrates that,
contrary to conventional superconductors, the superconductivity in
SC-FM heterostructures is effectively {\it enhanced} by an applied
magnetic field.

The authors acknowledge D.~Vodolazov for valuable discussions.
This work was supported by the Flemish Science Foundation
(FWO-Vl), The Belgian Science Policy, the University of Antwerp
(GOA), and the ESF programme ``VORTEX''.

\noindent{\it Note added in proof:} Very recently, Priour and
Fertig [Phys. Rev. Lett. {\bf 93}, 057003 (2004)] studied a
similar system where an extremely thin SC and MDs were in
immediate vicinity. This leads to much sharper magnetic field
profiles, enhancing disorder in the mixed state.


\begin{thebibliography}{9}

\bibitem{kosterlitz} J.M. Kosterlitz and D.J. Thouless, J. Phys.
C: Solid State Phys. {\bf 6}, 1181 (1973); M.R. Beasley, J.E.
Mooij, and T.P. Orlando, Phys. Rev. Lett. {\bf 42}, 1165 (1979).

\bibitem{chibotaru} L.F. Chibotaru {\it et al.}, Nature (London) {\bf 408}, 833 (2000); {\it
ibid.} Phys. Rev. Lett. {\bf 86}, 1323 (2001); V.R. Misko {\it et
al.}, Phys. Rev. Lett. {\bf 90}, 147003 (2003).

\bibitem{misko}  M.V. Milo\v{s}evi\'{c} and F.M. Peeters, Phys. Rev. B {\bf 68}, 024509 (2003).

\bibitem{bedanov} V.M. Bedanov and F.M. Peeters, Phys. Rev. B {\bf 49}, 2667 (1994).

\bibitem{schuller}  J.I.~Martin {\it et al.}, Phys. Rev. Lett. {\bf 79}, 1929 (1997).

\bibitem{schuller2}  J.I.~Martin {\it et al.}, Phys. Rev. Lett. {\bf 83}, 1022 (1999).

\bibitem{vanbael}  M.J. Van Bael {\it et al.}, Phys. Rev. B {\bf 59}, 14674 (1999).

\bibitem{schweigert1}  V.A.~Schweigert, F.M.~Peeters, and P.S.~Deo, Phys. Rev. Lett. {\bf 81}, 2783 (1998).

\bibitem{doria}  M.M.~Doria, J.E.~Gubernatis, and D.~Rainer, Phys. Rev. B
{\bf 39}, 9573 (1989).

\bibitem{lange}  M. Lange {\it et al.}, Phys. Rev. Lett. {\bf 90}, 197006 (2003).

\end{thebibliography}
\end{document}